\documentclass[a4paper,twocolumn,11pt,accepted=2025-03-31]{quantumarticle}
\pdfoutput=1
\usepackage[numbers]{natbib}
\usepackage[utf8]{inputenc}
\usepackage[english]{babel}
\usepackage[T1]{fontenc}
\usepackage{amsmath,amssymb}
\usepackage{hyperref}
\usepackage{graphicx}% Include figure files
\usepackage{dcolumn}% Align table columns on decimal point
\usepackage{bm}% bold math
\usepackage{xcolor}
\usepackage{tikz}
\usepackage{braket}
\usepackage[caption=false]{subfig}
\usepackage{stackengine}

\newcommand{\be}{\begin{equation}}
\newcommand{\ee}{\end{equation}}
\newcommand{\tr}{\text{Tr}}
\newcommand{\IABC}{I(A\!:\!B\!:\!C)}

\begin{document}

\title{Exact, Average, and Broken Symmetries in a Simple Adaptive Monitored Circuit}

\author{Zhi Li}
\affiliation{Perimeter Institute for Theoretical Physics, Waterloo, Ontario N2L 2Y5, Canada}
\email{zli@perimeterinstitute.ca}
% \orcid{0000-0002-2445-2701}
\author{Zhu-Xi Luo}
\email{zhuxi\_luo@g.harvard.edu}
% \orcid{0000-0003-1985-4623}
\affiliation{Department of Physics, Harvard University, Cambridge MA-02138, USA}
% \homepage{http://quantum-journal.org}
% \orcid{0000-0003-0290-4698}
% \thanks{You can use the \texttt{\textbackslash{}email}, \texttt{\textbackslash{}homepage}, and \texttt{\textbackslash{}thanks} commands to add additional information for the preceding \texttt{\textbackslash{}author}. If applicable, this can also be used to indicate that a work has previously been published in conference proceedings.}
% \author{Marcus Huber}
% \affiliation{Institute for Quantum Optics \& Quantum Information (IQOQI), Austrian Academy of Sciences, Boltzmanngasse 3, Vienna A-1090, Austria}
% \orcid{0000-0003-1985-4623}
% \author{Cassandra Granade}
% \affiliation{Microsoft Research, Quantum Architectures and Computation Group, Redmond, WA 98052, USA}
% \author{Johannes Jakob Meyer}
% \affiliation{Dahlem Center for Complex Quantum Systems, Freie Universität Berlin, 14195 Berlin, Germany}
% \orcid{0000-0003-1533-8015}
% \author{Victor V. Albert}
% \affiliation{Institute for Quantum Information and Matter \& Walter Burke Institute for Theoretical Physics, Caltech, Pasadena, CA 91125, USA}
% \orcid{0000-0002-0335-9508}
\maketitle

\begin{abstract}
Symmetry is a powerful tool for understanding phases of matter in equilibrium. Quantum circuits with measurements have recently emerged as a platform for novel states of matter intrinsically out of equilibrium. Can symmetry be used as an organizing principle for these novel states, their phases and phase transitions? In this work, we give an affirmative answer to this question in a simple adaptive monitored circuit, which hosts an ordering transition in addition to a separate entanglement transition, upon tuning a single parameter. Starting from a symmetry-breaking initial state, depending on the tuning parameter, the steady state could (i) remain symmetry-broken, (ii) exhibit the average symmetry in the ensemble of trajectories, or (iii) exhibit the exact symmetry for each trajectory. The ordering transition is mapped to the transition in a classical majority vote model, described by the Ising universality class, while the entanglement transition lies in the percolation class. Numerical simulations are further presented to support the analytical understandings.
\end{abstract}

\section{Introduction}

Exploring measurement-induced phases and phase transitions represents a frontier in our current understanding of non-equilibrium states of matter \cite{review}. Since the pioneering works \cite{SRN,Zeno,chan2019unitary} in 2018, the field has been vibrantly growing in the past few years \cite{gullans2020dynamical, li2019measurement, zabalo2020critical,PhysRevB.101.104302,PhysRevB.101.104301,gullans2020scalable,yoshida2021decoding,SangLiUltrafast}. 
While the earliest works focused on the competition between unitary gates and measurements, more variants have emerged. Of particular interest to us is the competition between non-commuting measurements in measurement-only circuits \cite{PhysRevResearch.2.023288,MeasurementOnly}, such as in the context of transverse field Ising \cite{lavasani2021measurement,SangTim,LangBuchler} and Kitaev spin liquid setups \cite{ZX,Adithya,zhu2023structured}, to mention a couple examples. One of the main exotic properties of these fruitful results is that the different ``phases'' of steady states are characterized not by conventional order parameters which break certain symmetries and are linear observable (of the form tr$(\rho \hat{O}))$. Instead, they are characterized by quantum information properties, typically the scaling of entanglement entropy. 

According to P.W. Anderson, physics is approximately the study of symmetry \cite{MoreIsDifferent}. It is natural to question whether symmetry plays a significant role in the study of monitored quantum circuits. The research community has made some attempts in this direction \cite{PhysRevB.101.104301,PhysRevB.101.104302,PRXQuantum.2.010352,PhysRevB.105.064306,PhysRevB.107.224303,PhysRevB.108.L041103,odea2022entanglement,PhysRevB.102.224311,PhysRevX.12.041002,PhysRevLett.129.120604,PhysRevLett.129.200602,PhysRevB.107.014308,PhysRevLett.131.020401,BAO2021168618}, 
such as: (1) The prototype measurement-induced phase transition arising from the competition between Haar random unitary gates and measurements can be formulated as the spontaneous breaking of not global, but replica symmetries \cite{PhysRevB.101.104301,PhysRevB.101.104302,PRXQuantum.2.010352}. 
(2) Symmetry of the circuit can determine the universality class of the measurement-induced phase transitions. For instance, in quantum automation circuits subject to measurements, the presence of $\mathbb{Z}_2$ symmetry leads to the parity-preserving class \cite{PhysRevB.105.064306,PhysRevB.108.L041103,odea2022entanglement}, while the absence of symmetry leads to the directed percolation class \cite{PhysRevB.102.224311}. 
(3) Symmetries can enrich the phase diagram of monitored circuits. 
For example, with $U(1)$ symmetry, there can be additional transitions in the volume-law phases \cite{PhysRevX.12.041002,PhysRevLett.129.120604,PhysRevLett.129.200602,PhysRevB.107.014308}. Symmetries imposed in the circuit can also be enlarged by dynamical replica symmetries, resulting in a variety of interesting phases and phase transitions \cite{BAO2021168618}. 
Additionally, references \cite{PhysRevB.107.214201,PhysRevB.108.214302,PhysRevB.109.125148} have examined the role of symmetries from an error correction perspective.

In this work, we will focus on the circuits with natural global symmetries, and try to interpret the known dynamical phases from a symmetry perspective. We examine a simple design of adaptive monitored circuits with only competing measurements and post-measurement feedbacks, both preserving a discrete $\mathbb{Z}_2$ Ising symmetry %$U=\prod_i X_i$. 
Our circuits are extensions of projective transverse Ising models \cite{SangTim,LangBuchler,lavasani2021measurement}. 
In addition to modifying the measurement schedule, we further add feedbacks to guide toward ferromagnetic outcomes. 

We find that the phase diagram is indeed nicely organized by symmetry, although besides the symmetry of pure state of each quantum trajectory: $U\ket{\psi}\propto\ket{\psi}$ (which we call exact symmetry), we also need to consider the symmetry of the ensemble $\rho$ incorporating the random schedule of measurements and their outcomes: $U\rho U^{-1}=\rho$ (which is called average symmetry). 
The idea is related to the recent explorations of the condensed matter community on decoherence and disorders in the contexts of average symmetry-protected topological orders \cite{Ma:2022pvq,Zhang:2022jul,ma2023topological,Lee:2022hog}, mixed state topological phases \cite{PhysRevB.95.075106}, average symmetries and their anomalies \cite{hsin2023anomalies,zhou2023reviving,zang2023detecting}.

\begin{figure}[htbp]
\centering
\begin{tikzpicture}[scale=1.5]
\draw[->,very thick] (0,0)--(4,0);
\filldraw[blue] (1.7,0) circle (1.5pt);
\node at (4.2,0) {$p$};
\node at (1.7,-0.25) {\textcolor{blue}{$p_E$}};
\node at (-0.5,0.25) {$(1+1)d$};
\node at (0.7,0.25) {exact PM};
\node at (0.7,-0.25) {exact sym.};
\node at (2.7,0.25) {average PM};
\node at (2.7,-0.25) {average sym.};
\filldraw[blue] (3.85,0) circle (1.5pt);
\node at (3.9,0.25) {\textcolor{blue}{FM}};
\node at (3.9,-0.25) {\textcolor{blue}{$p_O=1$}};
\end{tikzpicture}
\\
\vspace{1cm}
\begin{tikzpicture}[scale=1.5]
\draw[->,very thick] (0,0)--(4,0);
\filldraw[blue] (1.6,0) circle (1.5pt);
\node at (4.2,0) {$p$};
\node at (1.6,-0.25) {\textcolor{blue}{$p_E$}};
\node at (-0.5,0.25) {$(2+1)d$};
\node at (0.7,0.25) {exact PM};
\node at (0.7,-0.25) {exact sym.};
\node at (2.4,0.25) {average PM};
\node at (2.4,-0.25) {average sym.};
\filldraw[blue] (3.1,0) circle (1.5pt);
\node at (3.2,-0.25) {\textcolor{blue}{$p_O$}};
\node at (3.6,0.25) {FM};
\node at (3.75,-0.25) {no sym.};
\end{tikzpicture}
\caption{Sketches for phase diagrams. Top panel: (1+1)d. Bottom panel: (2+1)d. PM and FM stand for paramagnetic and ferromagnetic.}
\label{fig:sketch}
\end{figure}
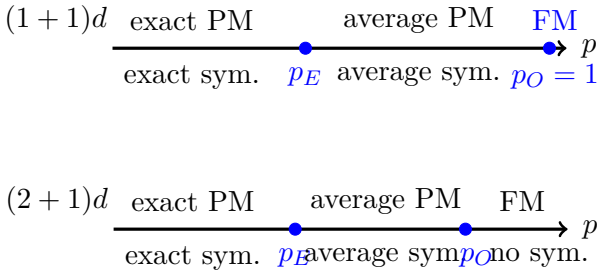

We sketch the phase diagrams of the (1+1)d and (2+1)d circuits in Fig.~\ref{fig:sketch}. 
Besides entanglement phase transitions $p_E$ in for both (1+1)d and (2+1)d circuits, there is an additional ordering transition $p_O$ for the (2+1)d circuit, which lies in the 2d classical Ising class. 
Before the entanglement transition $p<p_E$, the steady state of each trajectory exhibits exact $\mathbb{Z}_2$ symmetry. When $p_E<p<p_O$, each steady state no longer hosts the exact symmetry, but the ensemble of trajectories still exhibits the average $\mathbb{Z}_2$ symmetry. When $p>p_O$, the average symmetry is further broken, detectable by conventional linear order parameters such as the magnetization.

The remainder of the paper is organized as follows.  In Sec. \ref{sec:model} we describe the setup, followed by the discussion of the entanglement transition in Sec. \ref{sec:entangle}. The ordering transition is analyzed in Sec. \ref{sec:order}, where we map the circuit to a classical majority vote model.  
Then in the discussion section, we comment on the subtleties of initial states, the potential interpretation of spontaneous symmetry breaking, and directions for future research.

\section{Model and Terminology}
\label{sec:model}
We motivate our circuit model by the transverse field Ising model, schematically written as the following in arbitrary dimensions:
\begin{equation}
    H=-p\sum_{\langle ij\rangle} Z_i Z_j-(1-p)\sum_i X_i.
\end{equation}
The ground state is a ferromagnetic state $\otimes\ket{0}$ for $p=1$ and a paramagnetic state $\otimes\ket{+}$ for $p=0$.
It is well-known that there is an ordering phase transition at finite $p$, driven by the competition of $ZZ$ terms and $X$ terms.

\subsection{Model}
In a quantum circuit, we interpret $p$ and $1-p$ not as the coupling constant, but as the probability for the corresponding type of measurement to happen. To mimic the magnetic order and the ordering transition, one could consider two types of measurements that collapse the wavefunction into the $Z_iZ_j=1$ and $X_i=1$ subspaces respectively. 
However, postselections are not physical operations as the measurement outcomes are intrinsically random.
Still, one may attempt to enhance the $ZZ$ measurements and ``enforce" $ZZ=1$ as much as possible, by feedbacks that switch the eigenstates of $ZZ$.

More concretely, we start from a qubit chain with competing measurements and adaptive feedback \cite{buchhold2022revealing,PhysRevB.108.L041103,odea2022entanglement,PhysRevB.107.224303}. Periodic boundary condition is chosen for simplicity, and the initial state is ferromagnetically ordered  $\otimes\ket{0}$.
For each time step, we sequentially perform the following for each site $i$:
\begin{itemize}
    \item With probability $1-p$, we measure $X_i$.
    \item With probability $p$, we measure both $Z_{i-1}Z_{i}$ and $Z_{i}Z_{i+1}$.
    % for all nearest neighbour $j$ of $i$.
    Depending on the measurement outcomes, we perform the following feedback:
    \begin{itemize}
        \item[$\circ$] If $Z_{i-1}Z_i=Z_iZ_{i+1}=-1$, apply $X$ on site $i$;
        \item[$\circ$] If $Z_{i-1}Z_i=Z_iZ_{i+1}=+1$, do nothing; 
        \item[$\circ$] If $Z_{i-1}Z_i=-Z_iZ_{i+1}$, apply $X_i$ with probability $1/2$.
    \end{itemize} 
\end{itemize}
After the feedback, at least one of $Z_{i-1}Z_i$ and $Z_iZ_{i+1}$ will be $+1$. (In the 3rd situation, it is already the case even without feedback, but we choose to perform the random flip for analytical convenience.)
This updating rule is motivated by the classical majority-vote cellular automata \cite{PhysRevLett.55.657,book,Oliveira1992IsotropicMM,moore1997majority}, see also \cite{PhysRevLett.129.090404} for a quantum circuit model involving a similar updating rule.

It is evident that the dynamics (updating rules) has a $\mathbb{Z}_2$ Ising symmetry:
\begin{equation}\label{eq:IsingSymmetry}
    U=\prod_i X_i,
\end{equation}
while the initial state explicitly breaks this symmetry.
We deliberately choose this symmetry-broken initial state to highlight the restoration of exact or weak symmetries in the steady states. 

We will also consider a similar model for qubits arranged on a 2d square lattice. The setup is almost identical, except that for the $ZZ$-measuring steps we should measure all $Z_iZ_j$ where $j$ runs over all four nearest neighbors of $i$. The feedbacks are also designed to ensure most measurement outcomes are corrected to $+1$. Depending on the measurement outcomes, we apply $X_i$ (if there are more $-1$ than $+1$), apply $X_i$ with probability 1/2 (if there are two for each), or do nothing (if there are more $+1$ than $-1$).

\subsection{Exact and Average Symmetry}
The circuit has randomness coming from the measurement schedules (choices of the measurements) and the measurement outcomes.
In this work, we consider symmetries both of the trajectory -- the pure state given the measurement schedule and outcomes, and the ensemble -- the mixed state averaged over all measurement outcomes and/or measurement schedules. 

An exact symmetry is the normal symmetry defined for a pure state as $U|\psi\rangle = e^{i\theta} |\psi\rangle$. Here, we allow an arbitrary phase $\theta$ since a symmetry could act on the Hilbert space projectively. 
For a mixed state, one can define an average symmetry or a weak symmetry as $U \rho U^{-1} = \rho$. 
If $\rho$ is an ensemble of pure states $\{\ket{\psi_i}\}$, i.e.  $\rho=\sum p_i\ket{\psi_i}\bra{\psi_i}$, then the exact symmetry for each pure state $\ket{\psi_i}$ ensures the average symmetry for $\rho$. Note that for exact symmetry $\theta$ could vary for different $\ket{\psi_i}$, hence our exact symmetry is different from the ``strong symmetry" ($U\rho=e^{i\theta}\rho$) for mixed states used for example in Ref. \cite{Lee:2022hog}. 
In other words, exact symmetry is a symmetry defined for single trajectory, and is therefore a property of $\rho$ together with the decomposition $\rho=\sum p_i\ket{\psi_i}\bra{\psi_i}$, not just a property of $\rho$ alone. 
This is quite natural in our setting, where the quantum trajectories, determined by the realization of measurement schedule and measurement outcomes, provide a natural decomposition of $\rho$.

\section{Entanglement Transition}
\label{sec:entangle}

Our circuits are examples of quantum dynamics that fit into the stabilizer formalism \cite{gottesman1997stabilizer}.
Within this formalism, the Pauli feedback can only change the signs of the stabilizers.
Therefore, methods in analyzing measurement-only circuits and the entanglement transitions therein \cite{lavasani2021measurement,SangTim,LangBuchler} remain applicable here, provided we focus on structures and quantities that are invisible to the stabilizer signs.
Here, we extend the analysis in \cite{LangBuchler,lavasani2021measurement} to our modified setup %to adopt the symmetry-broken initial state, 
and revisit the transition from a symmetry perspective.

\subsection{Entanglement Dynamics and the Dual Percolation}

Let us first consider some simple examples to get some intuition on the entanglement dynamics.
Initially, the state is a product of $Z$-eigenstates.
As the circuit evolves, some $X$-eigenstates are generated by $X$ measurements. If a $ZZ$ measurement is applied on two $X$-eigenstates, then we get a two-qubit Bell state.

In general, at each moment, the system is always in a product state of some $Z$-eigenstates and some GHZ-like states $(\ket{s}\pm\ket{\bar{s}})/\sqrt{2}$ where $s$ is a classical bit string and $\bar{s}$ is its flip (for convenience, we regard an $X$-eigenstate as a GHZ state of size 1). 
Depending on whether a site appears in an $Z$-eigenstate or a GHZ-like state, we say the site lives in the background or in a GHZ cluster.
The dynamics for the stabilizer structure can be understood via the birth, split, merge, and death of the GHZ clusters:
\begin{itemize}
    \item  birth: measuring $X$ on a background site, a size-1 GHZ cluster is created;
    \item split: measuring $X$ in a size-$k$ GHZ cluster, the cluster splits into two clusters (size 1 and size $k-1$ respectively); 
    \item merge: measuring $Z_iZ_j$ where $i$ and $j$ belong to two different GHZ clusters, two clusters merge; 
    \item death: measuring $Z_iZ_j$ where one is in the background and one is in a GHZ cluster, the whole cluster disappears.
\end{itemize}

The above dynamics in $D$ spatial dimension can be captured by a bond percolation picture on $(D+1)$-dimensional hypercubic lattice.
At each time step, we activate a spatial $ij$ bond in the percolation picture iff $Z_iZ_j$ is measured in the circuit model, and we activate a temporal bond above site $i$ iff $X_i$ is \textit{not} measured.
The structure of the final state can be inferred from the percolation picture as follows: 
\begin{itemize}
\item sites that are connected (via activated bonds) to the initial slice are in the background and hence correspond to $Z$-eigenstates;
\item sites mutually connected but not connected to the initial slice are in a common GHZ cluster. 
\end{itemize}

\begin{figure}
      \begin{center}
\def\biga{\includegraphics[height=5.5cm]{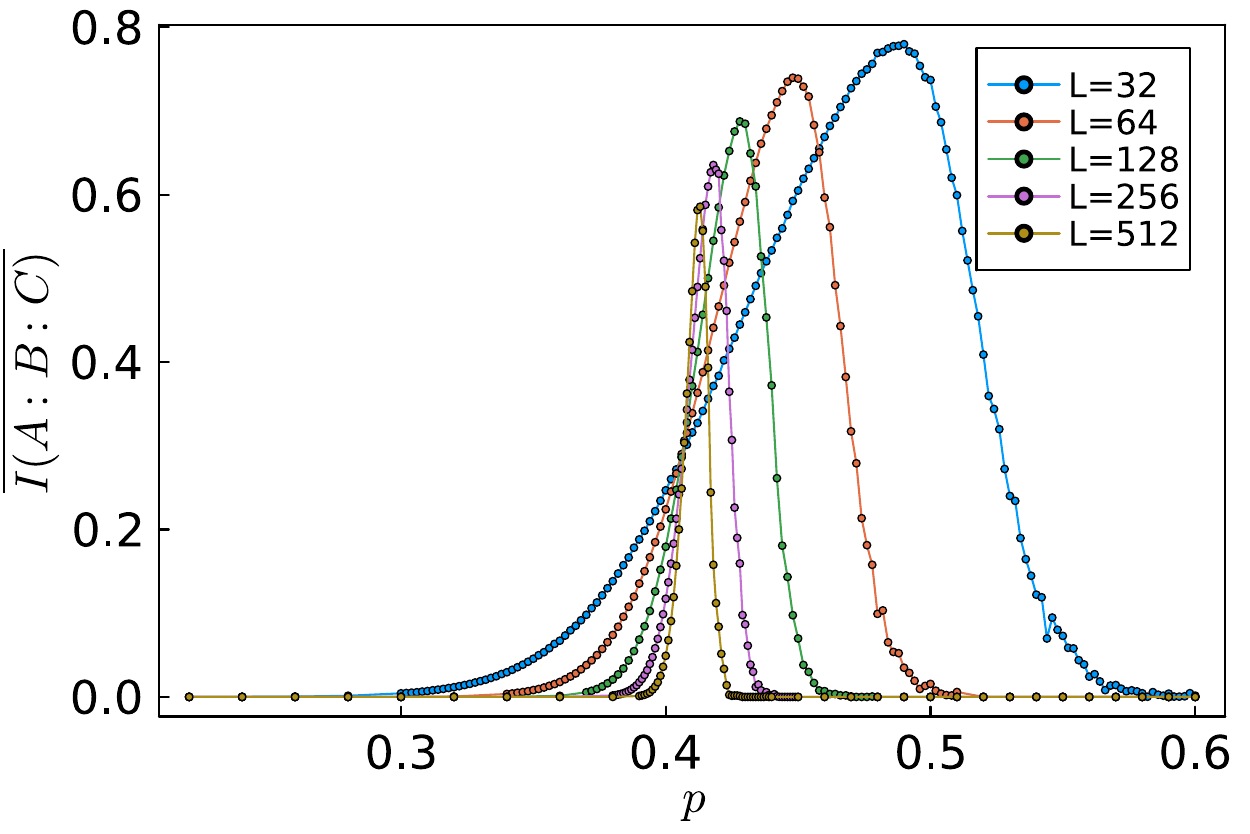}}
\def\inseta{\includegraphics[height=2.2cm]{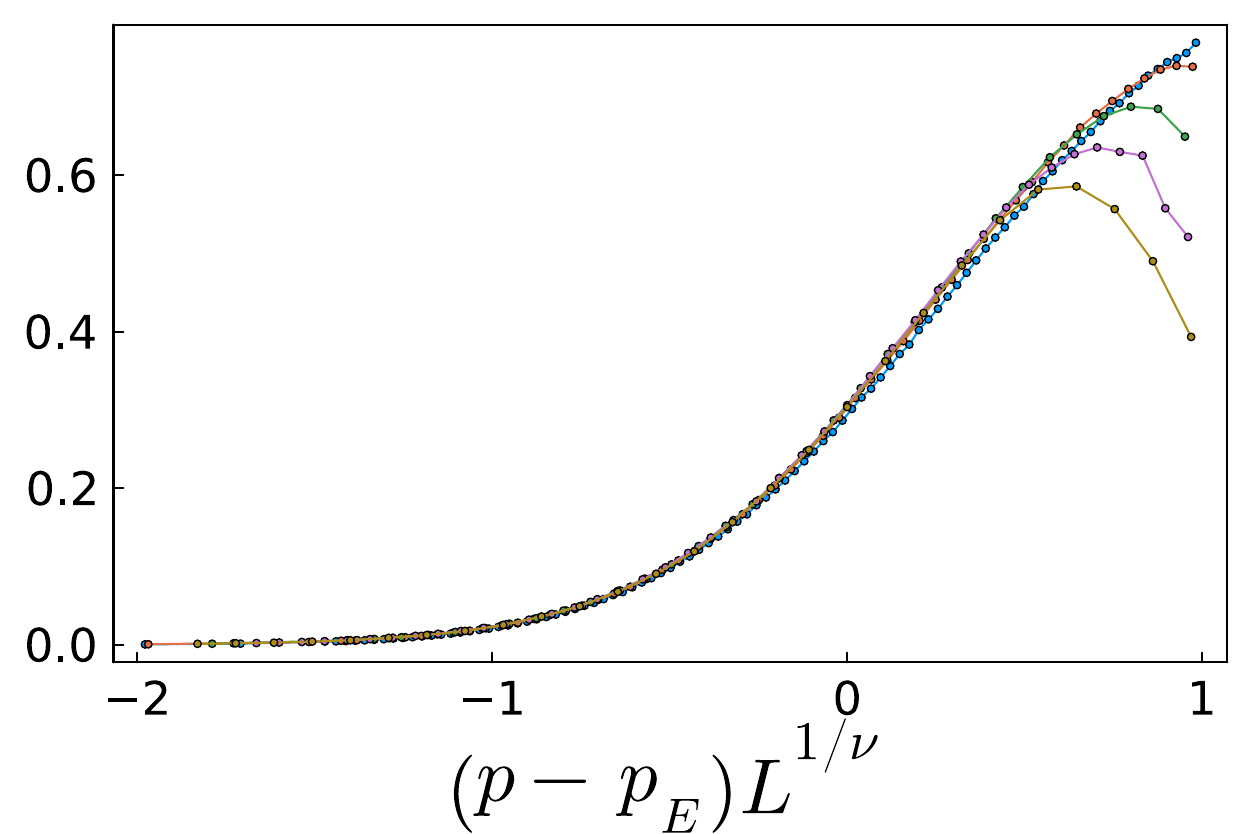}}
\def\insetb{\includegraphics[height=1.6cm]{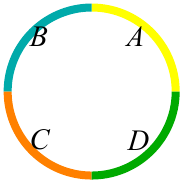}}
\stackinset{l}{1.4cm}{b}{30pt}{\insetb}{
\stackinset{l}{1.1cm}{t}{10pt}{\inseta}{\biga}
}
\def\bigb{\includegraphics[height=5.5cm]{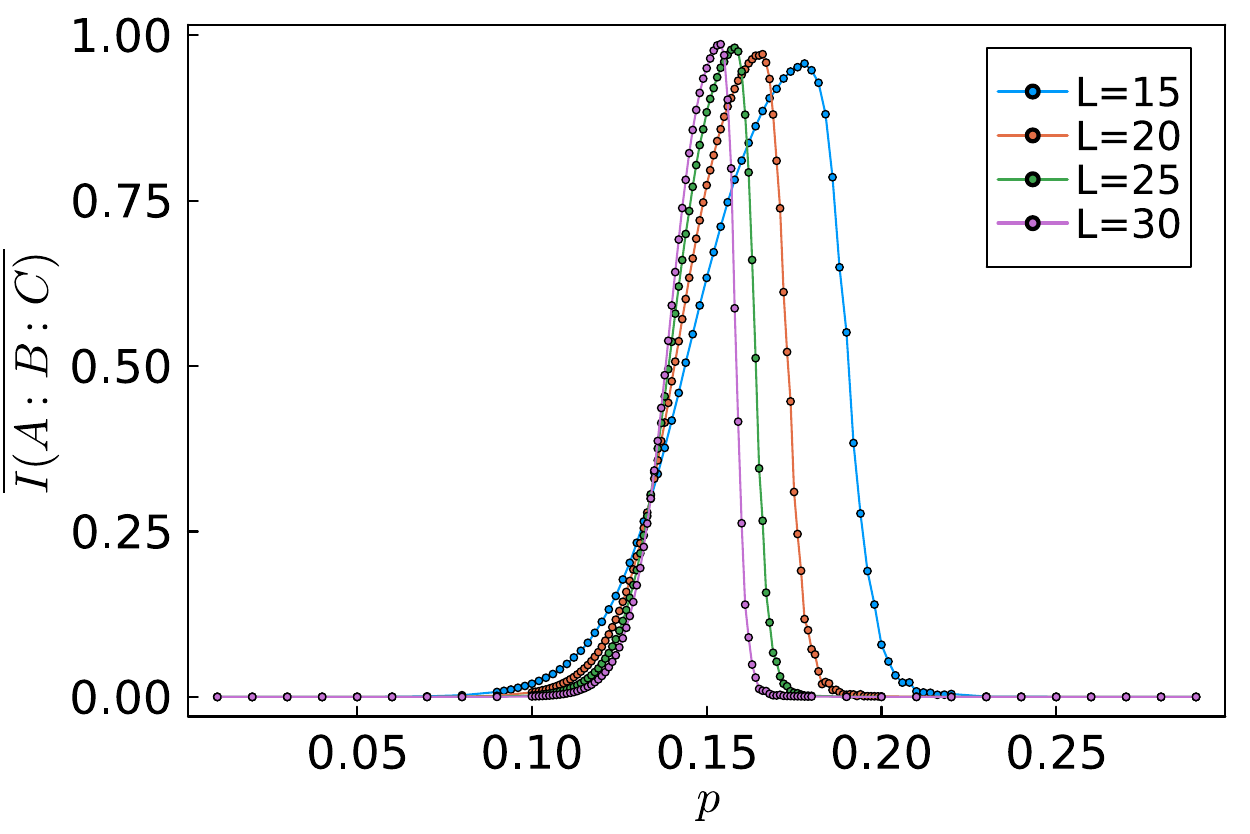}}
\def\insetc{\includegraphics[height=2cm]{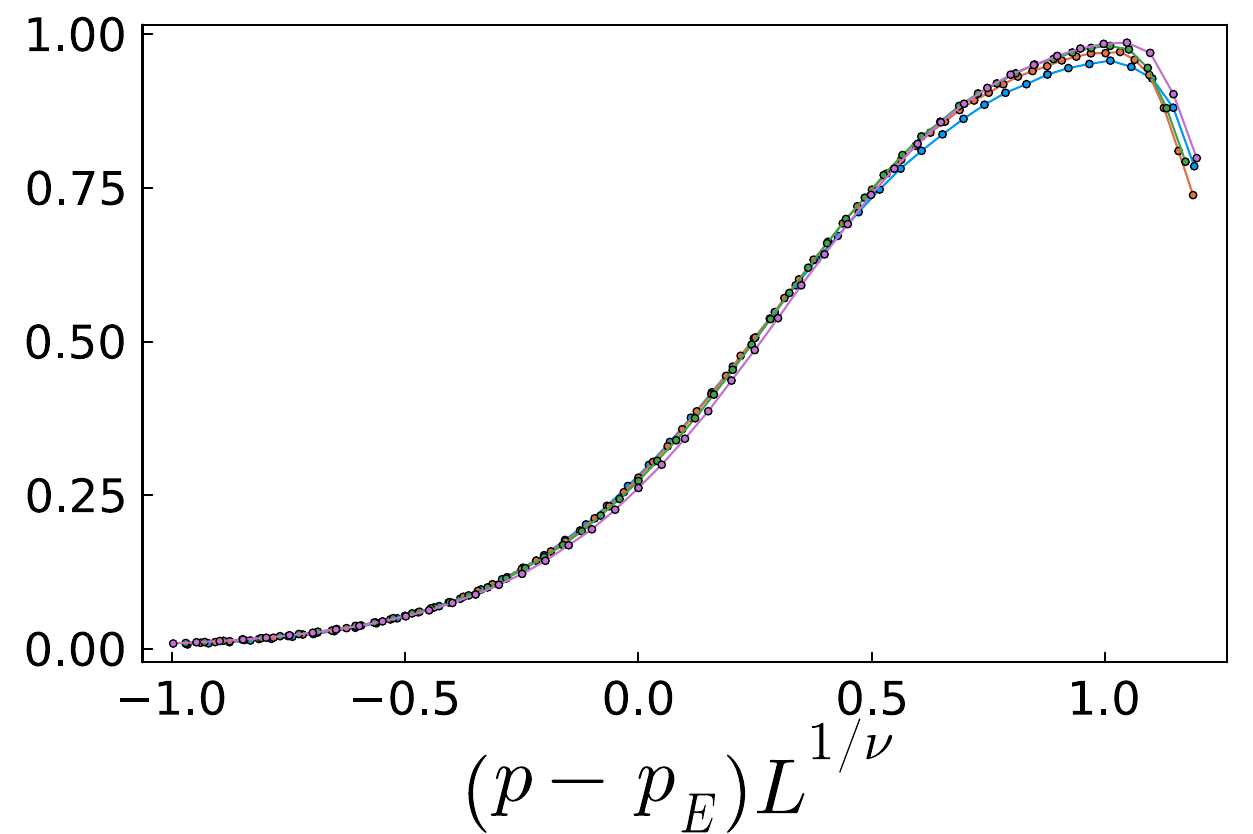}}
\def\insetd{\includegraphics[height=1.7cm]{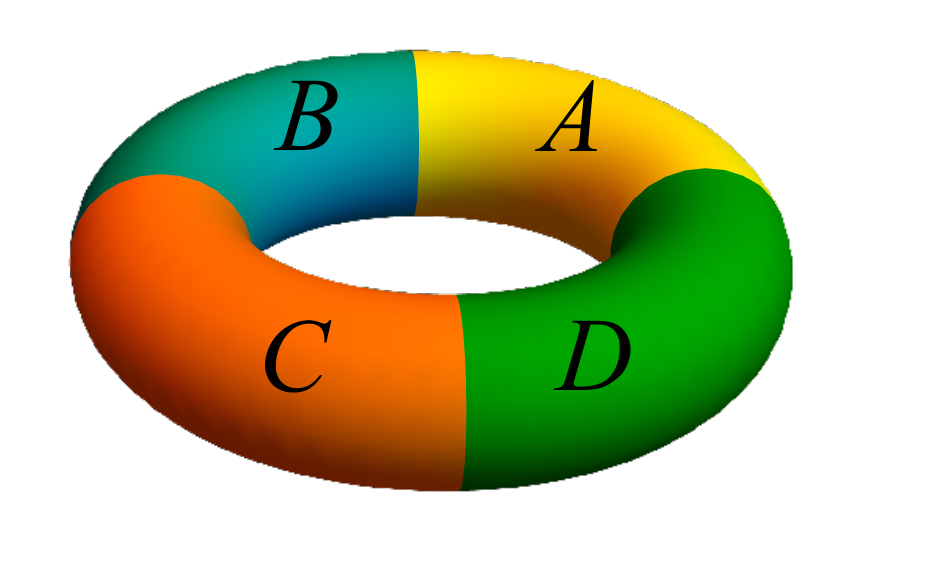}}
\stackinset{l}{1.4cm}{b}{30pt}{\insetd}{
\stackinset{l}{1.3cm}{t}{10pt}{\insetc}{\bigb}
}
\end{center}
    \caption{$\overline{I(A,B,C)}$ versus $p$ for (a) (1+1)d circuit and (b) (2+1)d circuit. The crossings correspond to the transition points: $p_E^{(1d)}\approx 0.40(7)$ and $p_E^{(2d)}\approx 0.13(3)$. The subfigures show data collapse for $p$ near $p_E$, for $\nu_E^{(1d)}=4/3$ and $\nu_E^{(2d)}=0.87$. }
    \label{fig:Sghz}
\end{figure}

This percolation model is not the usual isotropic bond percolation, since measurements in our quantum circuits are locally correlated.
Nevertheless, there remains a phase transition at finite probability $p_E$ that is in the same universality class, corresponding to a transition in the structure of steady states: 
the percolated (unpercolated) phase corresponds to the situation where $Z$-eigenstates exist (do not exist) in the final state.   
%\zhi{add below}
In the following, we still call it the entanglement transition, since (1) it has the same origin as the transition in measurement-only circuits, (2) the transition in the stabilizer structure, and (3) the fact that it can be probed by entanglement entropies such as the tripartite information:
\be
\begin{split}
\IABC=\ & S_A+S_B+S_C+S_{ABC}\\
& -S_{AB}-S_{BC}-S_{AC},
\end{split}
\label{eq:conditional}
\ee
(see Fig.~\ref{fig:Sghz} for the geometry).

$I(A\!:\!B\!:\!C)$ is related to the conditional mutual information $I(A,B|C)$ which in general detects ``large stabilizers" \cite{SangLiUltrafast}. 
In our model, it exactly measures the number of GHZ-clusters with nontrivial supports on all of the four regions $A$, $B$, $C$, $D$.
%\bookmark
In Fig.~\ref{fig:Sghz} we plot $\overline{\IABC}$ versus $p$ for the (1+1)d and (2+1)d circuits. 
It, as well as all subsequent numerics, is computed from averaging over different trajectories, indicated by the overline. 
At the transition, we do not see a step function behavior as in the case of symmetric initial states \cite{lavasani2021measurement}, which can be easily understood from the prototype wavefunctions $\otimes\ket{+}$ and $\otimes\ket{0}$ for two sides of the entanglement transition.
Instead, it shows a sharp peak: only near the percolation phase transition point can we have macroscopically connected clusters that are not connected to the initial time slice. The transition points are at $p_E^{(1d)}\approx 0.40(7)$ and $p_E^{(2d)}\approx 0.13(3)$ for (1+1)d and (2+1)d circuits. In Fig.~\ref{fig:Sghz} we plot the data collapse with the critical exponents $\nu_E^{(1d)}\approx 4/3$ and $\nu_E^{(2d)}\approx 0.87$, matching those for the 2d %\cite{kesten1980critical} 
and 3d \cite{PhysRevE.57.230,PhysRevE.87.052107,xu2014simultaneous} percolation universality classes.

\subsection{Exact v.s. Average Symmetry}
Importantly, a GHZ-like state is $\mathbb{Z}_2$ symmetric with $U$ in Eq.(\ref{eq:IsingSymmetry}):
\begin{equation}
    (\prod_i X_i)\ket{\text{GHZ}}=\pm\ket{\text{GHZ}},
\end{equation}
and a $Z$-eigenstate is not:
\begin{equation}
\braket{0|X|0}=\braket{1|X|1}=0.
\end{equation}
More generally, $\braket{U}=0$ if there exists a $Z$-eigenstate portion in the steady state; $\braket{U}=\pm 1$ if there does not.

Let us apply this observation in our circuits.
If $p<p_E$, namely, if the $X$ measurements dominate, there is no percolation hence all sites at the final slice belong to some GHZ clusters. Therefore, $\braket{U}=\pm 1$ for each trajectory. In other words, although the initial state explicitly breaks the symmetry, the symmetry is \emph{restored} by the quantum circuit to an \emph{exact} level: the final state of each quantum trajectory is $\mathbb Z_2$ symmetric.

On the other hand, if $p>p_E$, namely, if $ZZ$-bond measurements dominate, then percolation happens and there exist some $Z$-eigenstate factors in the final state. Therefore, $\braket{U}=0$. In this case, there is no exact symmetry at the trajectory level.
Nevertheless, we will show in Sec. \ref{sec:SB} that the $\mathbb Z_2$ symmetry is still restored as an \emph{average} symmetry at the level of density matrices, as long as $p$ is not too large. 
\begin{figure}
    \begin{center}
          \subfloat[]{\includegraphics[width=0.5\linewidth]{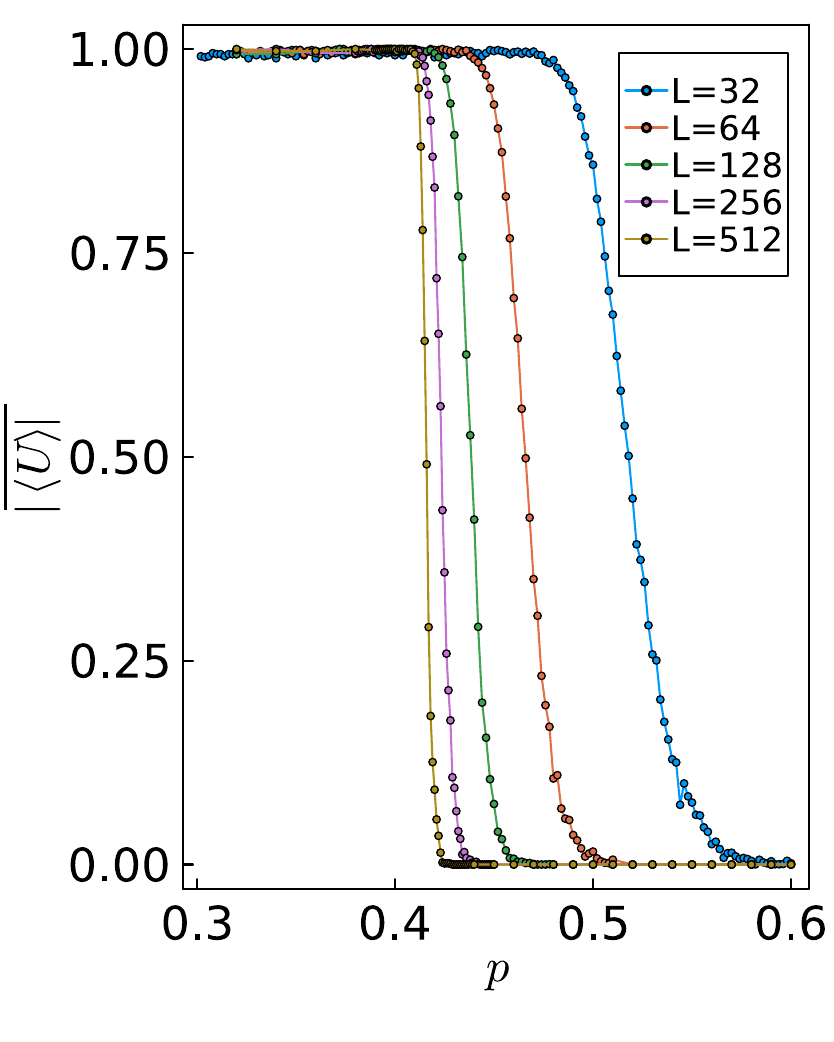}}
          \subfloat[]{\includegraphics[width=0.5\linewidth]{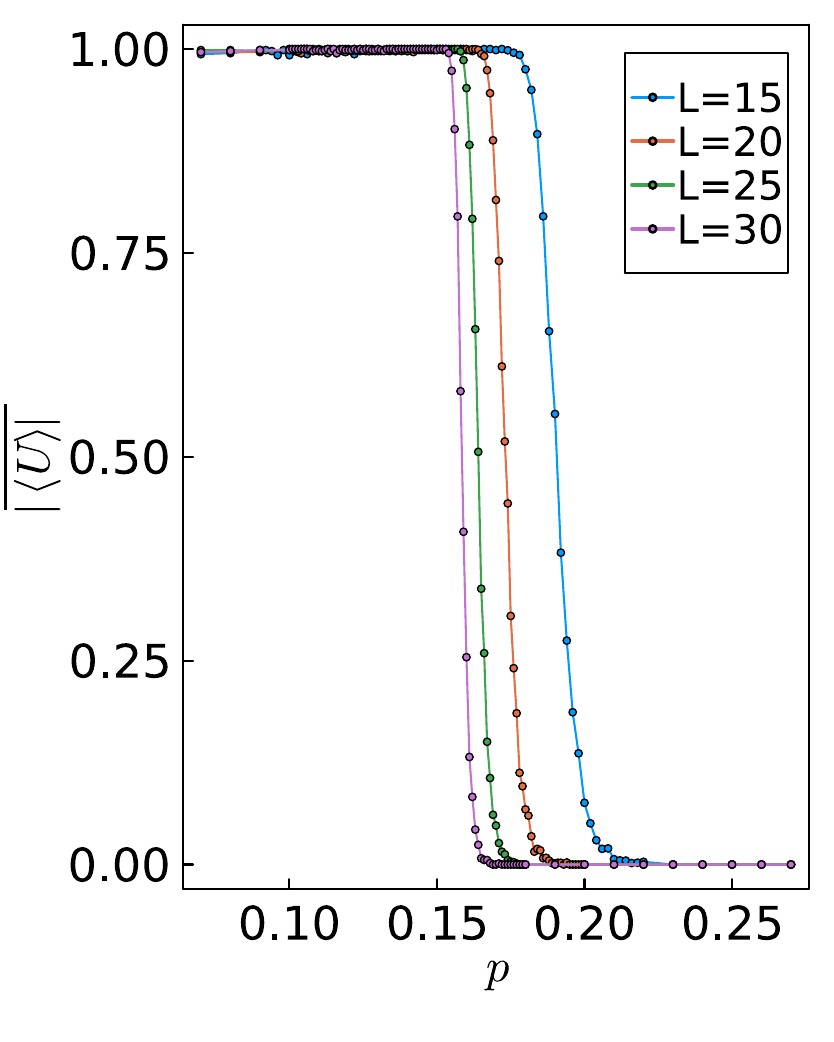}}
    \end{center}
\caption{$\overline{|\braket{U}|}$ versus $p$ for (a) (1+1)d circuit and (b) (2+1)d circuit.}
    \label{fig:PiX}
\end{figure}

In Fig.~\ref{fig:PiX} we plot $\overline{|\braket{U}|}$ versus $p$ for the (1+1)d and (2+1)d circuits. Note that we have to consider ${|\braket{U}|}$ instead of ${\braket{U}}$ since the sign of the latter is still random in the $X$ dominated phases. 
As expected, we see sharp step functions and the dropping point converges to the corresponding $p_E$, signaling a phase transition and the restoration of the exact symmetry in the $p<p_E$ side.

\section{Ordering Transition}
\label{sec:order}

In this section, we turn to the ordering transition that exists at spatial dimensions larger than one. We will present how this transition can be understood from a classical majority vote model in the same dimension.

\subsection{Classical Majority Vote Model}\label{sec:CMV}
\newcommand{\CMV}{\text{MV}$_c$}

We start by reviewing the classical majority vote (\CMV{}) model.
The \CMV{} model is usually formulated as follows.
There is one classical spin $\pm 1$ living on each lattice site.  At each time step, we pick a spin $i$ and check the signs of its nearest neighbors. 
\begin{itemize}
\item if its neighbor has a majority sign, then reset the spin $i$ to agree with the majority with probability $1-q$ and disagree with the majority with the probability $q$;
\item otherwise, randomly ($\pm 1$ with probability 1/2) reset the spin $i$. 
\end{itemize}
In this formulation, the ``out-of-majority" or noise parameter $q$ should satisfy $q<1/2$.
With the identification $q=\frac{1-p}{2}$, the above model can be equivalently described as follows:
\begin{itemize}
    \item with probability $1-p$, randomly reset spin $i$;
    \item with probability $p$, reset spin $i$ to agree with the majority sign (if exists) in its neighbor, or reset it randomly if there is no majority sign. 
\end{itemize}
The equivalence is evident: in the second formulation, the final probability for spin $i$ to agree with the majority is $p+\frac{1-p}{2}=1-q$ (if the majority sign exists), or the spin $i$ is reset randomly (if not), both in accordance with the rules in the first formulation.

The \CMV{} model has been extensively studied in the literature. In one spatial dimension, it is equivalent to the Glauber dynamics of the 1d Ising model (see for example Chapter 11 of \cite{book}), such that the distribution of classical configurations is the Boltzmann-Gibbs distribution. 
There is no transition for the \CMV{} model in 1d: an infinitesimal noise rate will drive the system into a disordered phase. 
In fact, it is quite difficult, and once thought impossible, for a 1d classical cellular automata to have a robust ordering against noise, consistent with the absence of long-range order at finite temperature in 1d statistical mechanics \cite{peierls_1936,PhysRev.136.A437,landau2013statistical}. Only delicately designed cellular automata can lead to a robust ordered phase in 1d \cite{GAC,Gac_guide}. 

On the other hand, the \CMV{} model in two and higher dimensions shows different behavior. 
Although they violate the detailed balance condition \cite{PhysRevLett.55.657,book} and avoid analytical solutions, it has been confirmed that there exists an order-disorder phase transition at finite $q$. Particularly in 2d, the transition is in the 2d classical Ising universality class \cite{Oliveira1992IsotropicMM}.

\subsection{Reduction to Classical Majority Vote}

Based on the discussions above, in the $ZZ$-dominant phase $p>p_E$, the steady states have entanglement structures that are in the same phase as the $Z$-basis product state.
With this observation in mind, let us approximate the steady state as an ensemble of classical states. Since an $X$ measurement results in the state $\ket{\pm}=(\ket{0}\pm\ket{1})/\sqrt{2}$, it is natural to view it as a noise in the classical approximation, which resets the bit into $\pm 1$ with half-half probability. 
Namely, the quantum circuit now reduces to the second formulation of the \CMV{} model with the same parameter $p$, with the same initial states.

Such reduction is certainly just an approximation at the trajectory level. However, in the appendix \ref{sec:proof}, we prove that it is exact at the ensemble level:
\be\label{eq:QequalsC}
\rho_q(t)=\rho_c(t).
\ee
Here $\rho_q$ and $\rho_c$ denote the density matrices in the quantum circuit model and the \CMV{} model respectively.
The relation continues to hold even if we fix the measurement schedule and only average over the measurement outcomes.

\subsection{Average v.s. Broken Symmetry}\label{sec:SB}

Given Eq.(\ref{eq:QequalsC}) and the established findings in the \CMV{} model, we deduce an order-disorder phase transition at finite $p_O$.
This transition is characterized by the breaking of the $\mathbb Z_2$ Ising symmetry at the ensemble level: $\rho_q(t)$ is ordered and average symmetric for $p<p_O$ given that $\rho_c(t)$ in the \CMV{} model is; while $\rho_q(t)$ is disordered and breaks the average symmetry for $p>p_O$. 
One can use a conventional order parameter, say, $\tr(\rho M)$ where $M=\sum_i Z_i$ to detect the symmetry breaking (recall that our initial state is deterministic). 
It can also be detected by the long-range correlator $\overline{\langle Z_1Z_{L/2}\rangle}$, which gains a finite value in the ordered phase and vanishes in the disordered phase.

\begin{figure}
    \begin{center}
    \def\big{\includegraphics[height=5.5cm]{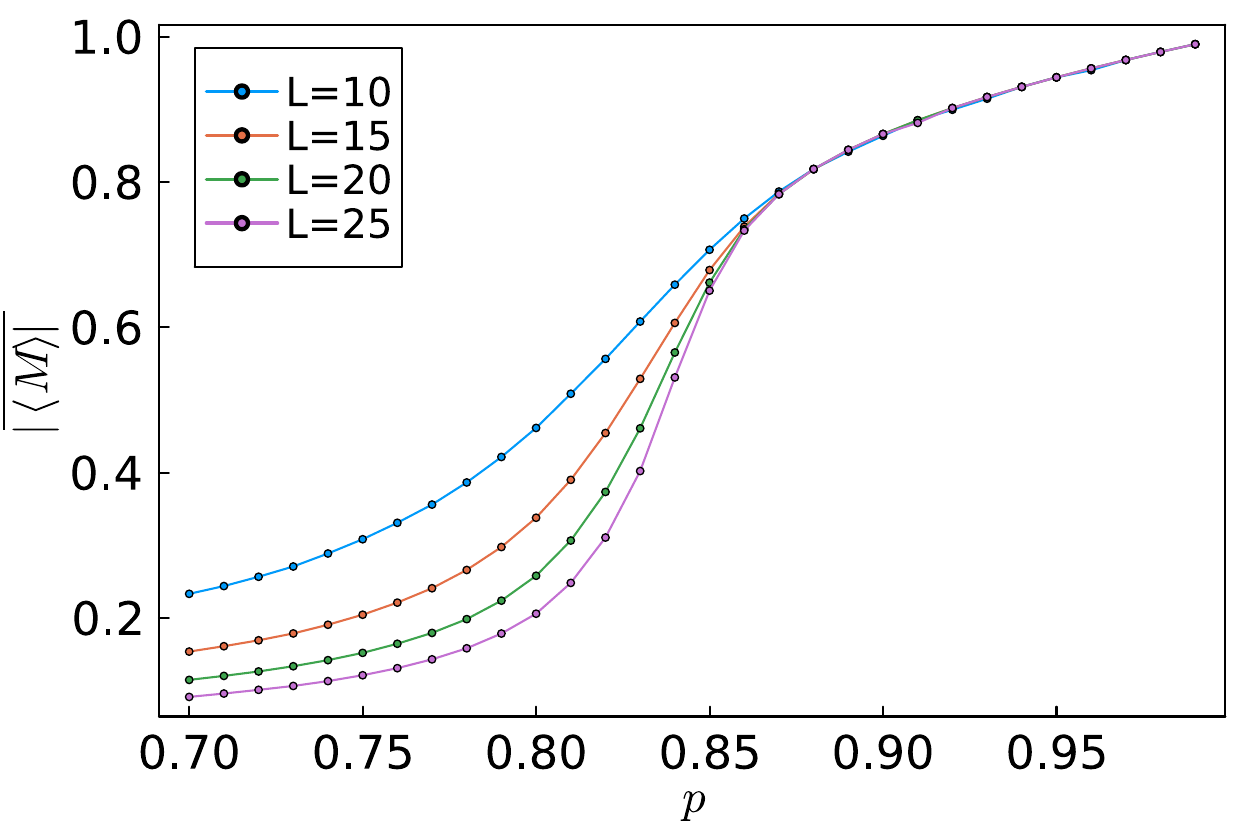}}
    \def\inset{\includegraphics[height=2.3cm]{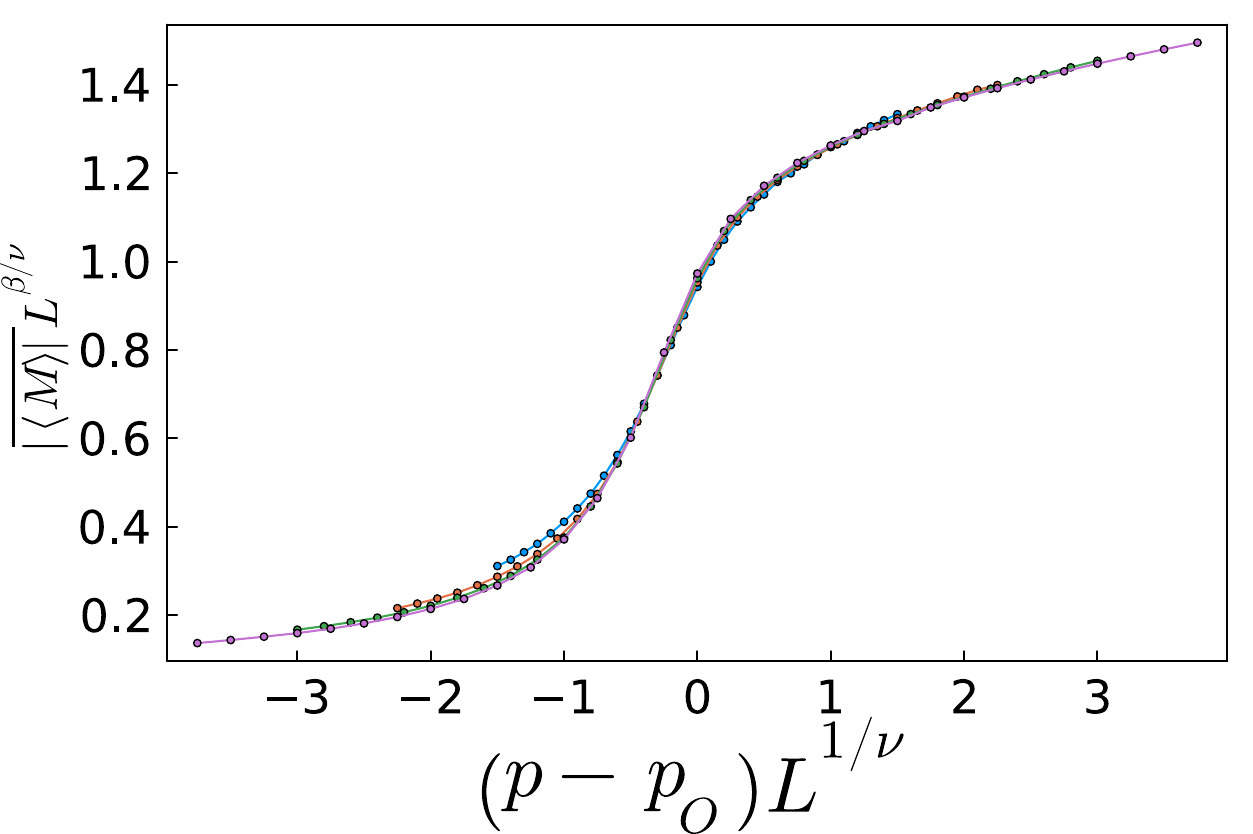}}
    \stackinset{r}{0.25cm}{b}{26pt}{\inset}{\big}
    \end{center}
    \caption{(a) $\overline{|\braket{M}|}$ versus $p$ in the (2+1)d circuit. The numerics are carried out using the stabilizer formalism \cite{gottesman1997stabilizer}.
    (b) Data collapse for $p$ near $p_O^{(2d)}\approx 0.85(0)$, with $\nu_O^{(2d)}\approx 1$ and $\beta_O^{(2d)}\approx 1/8$.}
    \label{fig:2D-Mag}
\end{figure}

Furthermore, we remark that the \CMV{} model provides a good picture for the quantum circuit for $p>p_E$ even at the trajectory level. As an illustrative example, the \CMV{} picture suggests that we can understand the steady states for $p_E<p<p_O$ as random product states in $Z$-basis (dressed by GHZ-clusters). Consequently, $\braket{M}$ should vanish for typical trajectories, not just for the ensemble $\rho_q$. 
In Fig.~\ref{fig:2D-Mag} we plot $\overline{|\braket{M}|}$ versus $p$ for the 2d circuit. 
We see a transition at $p_O\approx 0.85(0)$, in accordance with previous numerical finding $q\approx 0.075$ for the \CMV{} model \cite{Oliveira1992IsotropicMM} and the relation $q=\frac{1-p}{2}$.
${|\braket{M}|}$ indeed vanishes in the thermodynamics limit when $p<p_O$.
The critical exponents agree with the 2d Ising universality class, as shown by the data collapse with $\nu \approx 1,\ \beta \approx 1/8$ in the figure.

\section{Discussions}
\label{sec:discussion}

Although we have deliberately chosen the initial states to be a symmetry-broken state, we would like to point out that the symmetry breaking in the steady states may be understood as spontaneous (SSB). 

To understand that, let us first consider the usual spontaneous symmetry breaking in Hamiltonian systems.
Here, the ground state subspace is symmetric, yet physical states satisfying the cluster decomposition property are asymmetric.
For example, in the transverse field Ising chain, a GHZ state in the ferromagnetic phase is manifestly $\mathbb{Z}_2$ symmetric, but still in the symmetry broken phase with $\langle Z_i Z_j \rangle\neq 0$ for $i$ and $j$ that are far apart. 
Analogously, in the Monte-Carlo simulation of statistical mechanics, although Gibbs ensembles are always symmetric, the SSB phase does exist physically. 
Here, asymmetric external perturbations or random/asymmetric initial states are adopted in order to see the SSB.

In our case, the whole circuit that governs the dynamics preserves the Ising symmetry, but we have used an asymmetric initial state to detect the ability of symmetry breaking -- a property of the dynamics itself. 
The initial state does not have to be $\otimes\ket0$. 
In fact, a generalization of Eq.~\ref{eq:QequalsC} (see appendix) shows that a nonzero magnetization density in the initial state is enough to break the symmetry explicitly in the steady state for large $p$.

If, instead, the initial state is chosen to be $\otimes\ket{+}$ which is exactly symmetric, then the steady states will always preserve the exact symmetry, as observed in the projective transverse field model \cite{SangTim,LangBuchler}. In particular, $\braket{U}$ will always be 1 and the magnetization will always vanish for each trajectory.

However, even with the symmetric initial state $\otimes\ket{+}$, the symmetry breaking interpretation still holds: the manifestly symmetric state may live in a symmetry broken phase, in analogy to the Hamiltonian physics.
It can be made manifest in the language of mixed-state symmetries (see for example \cite{ma2023topological,PRXQuantum.4.030317,lessa2024strongtoweakspontaneoussymmetrybreaking,zhang2024strongtoweakspontaneousbreaking1form,liu2024diagnosingstrongtoweaksymmetrybreaking}).
If we assemble different realizations of the circuit into a diagonal density matrix, then all three phases have manifest strong Ising symmetry $\prod_i X_i \rho = \rho$ since both the initial state and the channel preserve the symmetry.
In the average paramagnet phase the Ising symmetry is broken from strong to weak, manifest from the finite order parameter of the strong Ising symmetry \cite{liu2024diagnosingstrongtoweaksymmetrybreaking} $\tr (\sqrt{\rho} Z_i Z_j \sqrt{\rho} Z_i Z_j)\approx 1$ 
when $i, j$ are well-separated, and the vanishing order parameter of the weak Ising symmetry $\tr (\rho Z_i Z_j)=0$. 
In the ferromagnetic phase, this weak symmetry is further  broken to trivial, as detectable from the finite order parameter of the weak symmetry $\tr (\rho Z_i Z_j)\approx 1$.
Regardless of the initial states, the symmetry properties in Fig.~\ref{fig:sketch} always holds: the strongly symmetric phase always has exact symmetry in each realization of the circuit; the strong-to-weak SSB phase preserves the weak or the average symmetry; and the ferromagnetic phase breaks the symmetry to trivial.
Therefore, regardless of the initial states, two phases transitions in Fig.~\ref{fig:sketch} can always be interpreted as symmetry-breaking phase transitions.
We point out that Ref. \cite{LangBuchler} also commented on the subtleties of initial states in monitored circuits and the possibility of obtaining ordering with feedback. However, the circuit carefully studied there was not adaptive and the symmetry perspective was not highlighted. 

We have focused on (1+1)d and (2+1)d circuits in this paper, yet our circuit setup as well as the analytical methods (the cluster model and the map to majority vote) are general and works for arbitrary dimensions and arbitrary lattices.  
Conversely, besides the majority vote, one could also consider other interesting classical dynamics. For example, it would be interesting to investigate the ``quantum version" of the nontrivial cellular automaton that shows robust order even in 1d \cite{GAC,Gac_guide}.

We hope that exact, (strong), average symmetries and their breaking can deepen our understanding of general non-equilibrium dynamics in quantum circuits, and again further exploration of this area is reserved for future studies.
For example, for a general symmetry group $G$, can there be (adaptive) monitored circuit that breaks the strong $G$ symmetry to a weak $H$ symmetry and further trivial symmetry? In the manuscript we have focused on the case where $H=G=\mathbb{Z}_2$ for the entanglement transition and $H=\{id\}$ for the ordering transition. 
It would be interesting to find the connections with other existing symmetric circuits such as \cite{PhysRevX.12.041002,PhysRevB.108.054307} and also other examples where $H\subset G$ is in general a subgroup of $G$. 
Furthermore, in the case where $G$ is anomalous, more exotic physics can potentially show up such as multipartite entanglement \cite{lessa2024mixedstatequantumanomalymultipartite} and generalized SPTs \cite{yu2025gaplesssymmetryprotectedtopologicalstates}.

\begin{acknowledgments}
We thank Ningping Cao, Meng Cheng, Maine Christos, Matthew Fisher, Timothy Hsieh, Peter Lunts, Henry Shackleton and Ruben Verresen for helpful discussions. 
Z.-X. L. thanks Zhen Bi, Carolyn Zhang and Jianhao Zhang for a related collaboration.
We thank anonymous referess for insightful comments and pointing out related references.
Z.L., via Perimeter Institute, is supported in part by the Government of Canada through the Department of Innovation, Science and Economic Development and by the Province of Ontario through the Ministry of Colleges and Universities. 
Z.-X.L. is supported by the Simons Collaboration on Ultra-Quantum Matter with Grant 651440 from the Simons Foundation.
\end{acknowledgments}

\bibliographystyle{quantum}

\bibliography{bib_quantum}

\onecolumn
\appendix

\appendix

\section{Reduction to Classical Majority Vote--the Proof}
\label{sec:proof}

In the quantum circuit model, there are two types of channels: (1) measure $X_i$ and average over the measurement outcomes, denoted by $\mathcal X_i$; (2) measure some $ZZ$ bond operators simultaneously around qubit $i$ and perform a Pauli unitary $X$ or $I$ according to the measure outcomes, denoted by $\mathcal F_i$.
Schematically, we have:
\begin{align}\label{eq:app-rho1}
    \Phi_q=\cdots\mathcal X_i\mathcal{F}_j\cdots.
\end{align}

In the classical majority vote model, writing it using quantum mechanics, there are also two types of channels: (1) reset a (qu)bit $i$ to up or down with half-half probability, denoted by $\mathcal{T}_i$; (2) same $\mathcal F_i$ as above.
Systematically, we have:
\begin{align}
\Phi_c=\cdots\mathcal{T}_{i}\mathcal{F}_j\cdots.
\end{align} 
We claim that, for initial state $\otimes\ket{0}$, we have:
\begin{equation}
    \rho_q=\rho_c.
\end{equation}

To prove it, we introduce a dephasing quantum channel $\mathcal{D}$ that convert a coherent wavefunction into a classical mixture of $Z$-basis states:
\begin{equation}
    \mathcal{D}=\otimes \mathcal{D}_i,
\end{equation}
where
\begin{align}
    \mathcal{D}_i(\rho)
    =\sum_{a=\{\pm 1\}} P_a\rho P_a=\sum_{a=\pm 1}(\frac{1+aZ_i}{2})\rho(\frac{1+aZ_i}{2}).
\end{align}
We will use the following two relations:
\begin{align}
&\mathcal X_i\mathcal{D}=\mathcal{D}\mathcal X_i=\mathcal{D}\mathcal{T}_i=\mathcal{T}_i,\label{eq:relation1}\\
&\mathcal{D}\mathcal{F}_i=\mathcal{F}_i\mathcal{D}.\label{eq:relation2}
\end{align}
With them in mind, the proof is then straightforward.
Since $\mathcal D(\rho(0))=\rho(0)$, we can insert a $\mathcal{D}$ before $\rho(0)$ in Eq.(\ref{eq:app-rho1}) and keep moving $\mathcal{D}$ to the left end using Eqs.(\ref{eq:relation1},\ref{eq:relation2}).
We get:
\begin{equation}
    \begin{aligned}
    \rho_q(t)
    &=\cdots\mathcal X_i\mathcal{F}_j\cdots\mathcal{D}(\rho(0))\\
    &=\mathcal{D}\cdots\mathcal T_i\mathcal{F}_j\cdots(\rho(0))\\
    &=\mathcal{D}\rho_c(t)\\
    &=\rho_c(t).
\end{aligned}
\end{equation}
The last equation is because $\rho_c(t)$ is already classical (diagonal in $Z$-basis).

Slightly more generally, we have
\begin{equation}
    \mathcal{D}\Phi_q=\Phi_q\mathcal{D}=\Phi_c\mathcal{D}.
\end{equation}
The proof is similar.

It remains to verify the two relations in Eqs.(\ref{eq:relation1},\ref{eq:relation2}).
For the first, it is enough to consider a single qubit:
\begin{equation}
    \sum_{a,b=\pm 1}\frac{1+aX}{2}\frac{1+bZ}{2}\rho\frac{1+bZ}{2}\frac{1+aX}{2}
    =\frac{1}{2}I.
\end{equation}
The second is because $\mathcal{F}$ only cares about and operates on the $Z$-basis information. Formally, we write $\mathcal F$ in the Kraus form $\mathcal F(\rho)=\sum_K K\rho K^\dagger$ with the Kraus operators $K=X\prod\frac{1\pm ZZ}{2}$ or $K=\prod\frac{1\pm ZZ}{2}$, and then notice that
\begin{equation}
\begin{aligned}
   \mathcal D\mathcal F(\rho)
   &=\sum_{a=\{\pm 1\}^n}\sum_K P_a K\rho K^\dagger P_a\\
   &=\sum_{a=\{\pm 1\}^n}\sum_K KP_{a'}\rho P_{a'} K^\dagger\\
 &=\sum_K\sum_{a'=\{\pm 1\}^n} KP_{a'}\rho P_{a'} K^\dagger\\
 &=\mathcal F\mathcal D(\rho).
\end{aligned}
\end{equation}
Here $a'$ is determined by the (anti-)commutation relation between $P_a$ and $K$, but it is enough to know that $a'$ also runs over $\{\pm 1\}^n$.

\end{document}